# Estimation of the Randomness of Continuous and Discrete Signals Using the Disentropy of the Autocorrelation

R. V. Ramos

rubens.ramos@ufc.br

*Lab. of Quantum Information Technology, Department of Teleinformatic Engineering – Federal University of Ceara - DETI/UFC, C.P. 6007 – Campus do Pici - 60455-970 Fortaleza-Ce, Brazil.*

*Abstract* — The amount of randomness in a signal generated by physical or non-physical process can reveal important information about that process. For example, the presence of randomness in ECG signals may indicate a cardiac disease. On the hand, the lack of randomness in a speech signal may indicate the speaker is a machine. Hence, to quantify the amount of randomness in a signal is an important task in many different areas. In this direction, the present work proposes to use the disentropy of the autocorrelation function as a measure of randomness. Examples using noisy and chaotic signals are shown.

## I. Introduction

A basic task in modern science is the data analysis. Medical, astronomical, economic, seismological, among many others types of data, generated by physical or non-physical processes, have to be analysed every day. In all cases these data are (intentionally or not) contaminated with noise. This noise is the source of randomness in the data that, depending on the goal desired can be useful or not. The randomness in meteorological data can be very harmful since it makes harder the weather forecast. On the other hand, the randomness in a cryptographic key is highly desired. Thus, to study the amount of randomness in a (continuous or discrete) signal is an important task since it can inform, for example, the presence of a disease in a ECG signal, an illegal operation in the stock market, to denounce a fake image or speech or, simply, to show the data considered is useless, like a cryptographic key with low randomness.

An important mathematical tool in the analysis of random signals is the autocorrelation function. For a continuous signal $s(t)$, the autocorrelation function is defined as

$$R(\tau) = \int_{-\infty}^{\infty} s(t) s^*(t-\tau) dt. \qquad (1)$$

Similarly, the autocorrelation at lag $k$ of the discrete signal $s_t$ is defined as

$$r_k = \frac{E\left[(s_t - \bar{s})(s_{t+k} - \bar{s})\right]}{\sigma_s^2} = \frac{1}{N} \frac{\sum_{t=1}^{N-k}(s_t - \bar{s})(s_{t+k} - \bar{s})}{\frac{1}{N}\sum_{t=1}^{N}(s_t - \bar{s})^2}, \qquad (2)$$

where $\bar{s}$ and $\sigma_s^2$ are, respectively, the mean value and variance of $s_t$. The symbol $E$ denotes the expected value. Since the autocorrelation function shows the similarity between a function and its delayed version, thereby showing the degree by which its value at one time is similar to its value at a certain later time, it is natural to use it to analyse the randomness contained in a signal: the larger the randomness, the smaller the similarity. However, the autocorrelation is a function and not a variable. Thus, in order to get a randomness measure based on the autocorrelation, one needs a function that maps the autocorrelation function to a real number. In this work, the disentropy is used for this task. As it will be show latter, the disentropy is a good candidate once it has a physical meaning, it can be easily calculated and unlike entropy its argument can be negative. The latter is very important because the autocorrelation function usually has negative values.

The present work is outlined as follows: In Section II the deformed Lambert functions and their associated disentropies are briefly reviewed; In Section III the disentropy of the autocorrelation function of some random signals is considered; Section IV shows the disentropy of the autocorrelation of a chaotic signals; At last, the conclusions are drawn in Section V.

## II. The Lambert-Tsallis $W_q$ Function and the Disentropy

The Lambert $W$ function is an important elementary mathematical function that has been used in different areas of mathematics, computer Science and physics [1-6]. Basically, the Lambert $W$ function is defined as the solution of the equation

$$W(z) e^{W(z)} = z. \qquad (3)$$

Equation (3) has infinite solutions, however, only two of them return a real value when the argument $z$ is real. In the interval $-1/e \leq z \leq 0$ there exist two real values of $W(z)$. The branch for which $W(z) \geq -1$ is the principal branch named $W_0(z)$ while the branch satisfying $W(z) \leq -1$ is named $W_{-1}(z)$. For $z \geq 0$ only $W_0(z)$ is real and for $z < -1/e$ there are not real solutions. The point $(z_b = -1/e, W(z_b) = -1)$ is the branch point where the solutions $W_0$ and $W_{-1}$ have the same value and $dW/dz = \infty$. The plot of $W(z)$ versus $z$ is shown in Fig. 1.

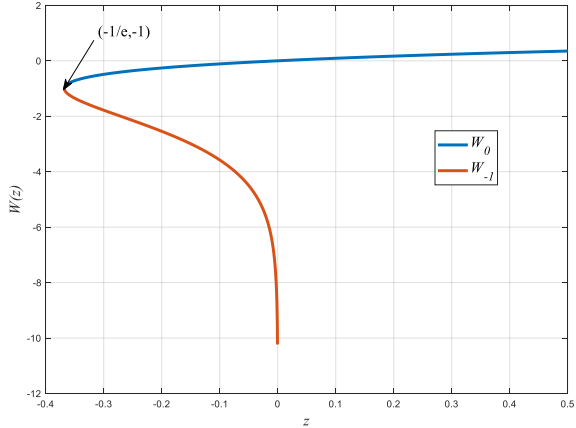

Fig. 1 – $W(z)$ versus $z$

Taking the logarithm in both sides of eq. (3) one gets

$$\log(z) = W(z) + \log[W(z)]. \tag{4}$$

Thereby, the entropy can be written as

$$S = \sum_i p_i \log(p_i) = \sum_i p_i W(p_i) + \sum_i p_i \log[W(p_i)]. \tag{5}$$

where $\{p_1, p_2, \ldots, p_n\}$ is a probability distribution. The term

$$D = \sum_i p_i W(p_i) \tag{6}$$

is named disentropy [7]. When the disentropy is minimal entropy is maximal and vice-versa. Equation (6) is the disentropy related to Boltzmann-Gibbs entropy. If eq. (3) is modified to

$$R_2(z) 2^{R_2(z)} = z, \tag{7}$$

the solution of eq. (7) is

$$R_2(z) = \log_2(e) W\left(\frac{z}{\log_2(e)}\right) \tag{8}$$

and the disentropy related to Shannon entropy is

$$D = \sum_i p_i R_2(p_i). \tag{9}$$

One can modify eq. (3) once more by exchanging the exponential by Tsallis $q$-exponential [7]

$$W_q(z) e_q^{W_q(z)} = z, \tag{10}$$

where Tsallis $q$-exponential is given by

$$e_q^z = \begin{cases} e^z & q = 1 \\ [1+(1-q)z]^{1/(1-q)} & q \neq 1 \ \& \ 1+(1-q)z \geq 0 \\ 0 & q \neq 1 \ \& \ 1+(1-q)z < 0 \end{cases}. \tag{11}$$

The function $W_q(z)$ is named Lambert-Tsallis function and it also has two branches with real solutions. Taking the $q$-logarithm in both sides of (10), one gets

$$\log_q(z) = W_q(z) + \log_q[W_q(z)] + (1-q) W_q(z) \log_q[W_q(z)]. \tag{12}$$

where

$$\log_q(z) = \begin{cases} \log(z) & x > 0 \ \& \ q = 1 \\ \dfrac{x^{(1-q)} - 1}{1-q} & x > 0 \ \& \ q \neq 1 \\ \text{undefined} & x \leq 0 \end{cases}. \tag{13}$$

Therefore, Tsallis $q$-entropy [8] can be written as

$$S_T = \sum_i p_i^q \log_q(p_i) = \sum_i p_i^q W_q(p_i) + \sum_i p_i^q \log_q[W_q(p_i)] + (1-q) \sum_i p_i^q W_q(p_i) \log_q[W_q(p_i)]. \tag{14}$$

The term

$$D_q = \sum_i p_i^q W_q(p_i) \tag{15}$$

is the disentropy related to Tsallis $q$-entropy.

At last, one can also modify eq. (3) by using the $q$-product operation [9]

$$R_{qQ}(z) \times_Q e_q^{R_{qQ}(z)} = z, \qquad (16)$$

where

$$a \times_q b = \max\left\{\left[a^{(1-q)} + b^{(1-q)} - 1\right]^{1/(1-q)}, 0\right\}. \qquad (17)$$

Doing $q = Q$ and taking the $q$-logarithm in both sides of (16) one gets

$$\log_q(z) = R_{qq}(z) + \log_q\left[R_{qq}(z)\right]. \qquad (18)$$

Thus, Tsallis $q$-entropy can now be written as

$$S_T = \sum_i p_i^q \log_q(p_i) = \sum_i p_i^q R_{qq}(p_i) + \sum_i p_i^q \log_q\left[R_{qq}(p_i)\right]. \qquad (19)$$

The term

$$D_{qq} = \sum_i p_i^q R_{qq}(p_i) \qquad (20)$$

is a second disentropy related to Tsallis $q$-entropy.

Some examples (upper branch) for $W_q$ and $R_{qq}$ are shown below [7,9]:

$$W_2(z) = \frac{z}{z+1}, \quad z > -1, \qquad (21)$$

$$W_{3/2}^+(z) = \frac{2(z+1) + 2\sqrt{2z+1}}{z}, \quad z > -1/2, \qquad (22)$$

$$R_{2,2}(z) = -\frac{1}{2z} \pm \frac{1}{2}\sqrt{\frac{1}{z^2} + 4}, \quad z \geq 0, \qquad (23)$$

$$R_{1/2,1/2}(z) = 2\left(z^{1/2} + 1\right) - 2\sqrt{2z^{1/2} + 1}, \quad z \geq 0. \qquad (24)$$

It can be shown the branch point of the Lambert-Tsallis $W_q$ function is $(z_b = \exp_q(1/(q-2))/(q-2), W_q(z_b) = 1/(q-2))$, for $q \neq 2$. There is no branch point with finite $z_b$ for $q = 2$. For $q = 1$, the branch point of Lambert $W$ function is recovered. The solution in the interval $z_b \leq z < 0$ is $W_q^-(z)$ while the solution in the interval $z_b \leq z < \infty$ is $W_q^+(z)$. For example, Fig. 2 shows the plot of $W_{q=3/2}$ versus $z$.

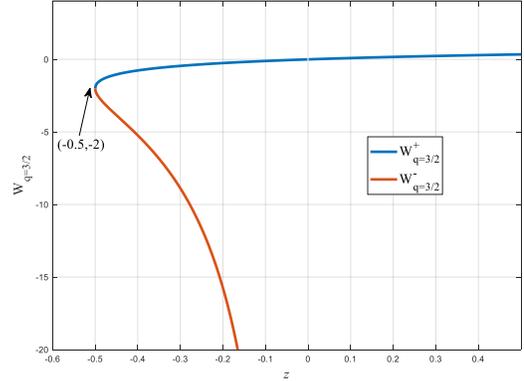

Fig.2. $W_{q=3/2}(z)$ versus $z$.

One may note it is also possible to modify eq. (3) by using the Kaniadakis $\kappa$-exponential [10]. In this case, the solutions will be the Lambert-Kaniadakis $W_\kappa$ functions, and another disentropy based on $W_\kappa$ can be introduced [11].

The disentropy can be used everywhere entropy is used [12], however, there are situations where only the disentropy can be used, like the quantumness measure proposed in [13] and the randomness measure proposed in this work. At last, although only the discrete disentropies were described, their continuous version is straightforward. For example, if $p(x)$ is a probability density distribution, then its disentropy $D_q$ is given by

$$D_q = \int_{-\infty}^{\infty} p^q(x) W_q(p(x)) dx. \qquad (25)$$

### III. THE DISENTROPY OF THE AUTOCORRELATION FUNCTION OF NOISY SIGNALS

In this work only the disentropy $D_{q=2}$ (eqs. (15) and (21)) will be used, since it can be easily calculated and the autocorrelation function is inside of its domain. Thus, the randomness measure of a continuous signal $s(t)$ is

$$D_2 = \int_{-\infty}^{\infty} \frac{R^3(\tau)}{R(\tau) + 1} d\tau. \qquad (26)$$

while for a discrete series $s_t$ one gets

$$D_2 = \sum_{n=1}^{N} \frac{r_n^3}{r_n + 1}, \qquad (27)$$

where $r_n$ is the $n$-th value of the discrete autocorrelation function of $s_t$.

Let us start using the signal given in eq. (28).

$$s(t) = \sin(2\pi t) + 3\sin(3\pi t) + A\varepsilon(t). \qquad (28)$$

In eq. (28) $A$ is the variable that controls the noise power (the noise power is proportional to $A^2$) and $\varepsilon(t)$ is a random variable with uniform distribution in the interval [0,1]. The curve disentropy versus $A$ for $s(t)$ is shown in Fig. 3.

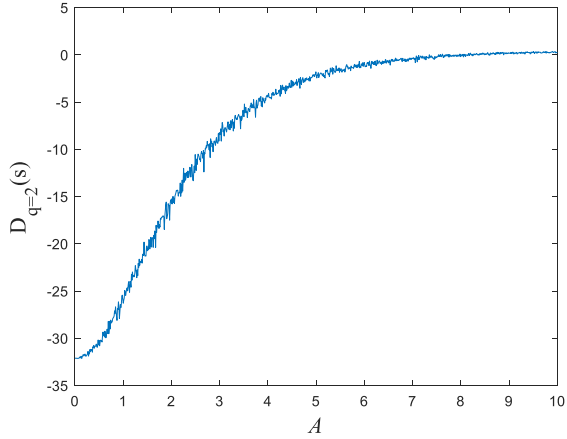

Fig. 3 – $D_2(s)$ versus $A$ for $s(t)$ given in eq. (28).

For each value of $A$, eq. (28) was simulated only once, this is the reason the curve is not so smooth. However, its behaviour is very clear, the larger the noisy power the larger the randomness, until the saturation is reached.

The second signal to be considered is given in eq. (29), a signal with noise in the frequency.

$$s(t) = 5\sin\left(2\pi\left[5 + A\varepsilon(t)\right]t\right). \qquad (29)$$

The curve disentropy versus $A$ for $s(t)$ is shown in Fig. 4.

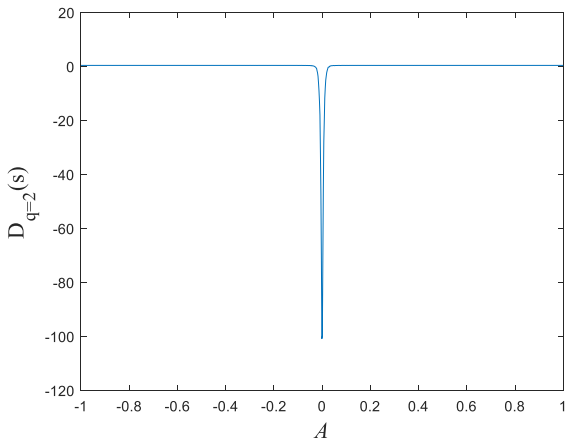

Fig. 4 – $D_2(s)$ versus $A$ for $s(t)$ given in eq. (29).

As one can see in Fig. 4, once more the larger the noise power the larger the randomness.

At last, eq. (30) shows a signal with phase noise.

$$s(t) = 5\sin\left(2\pi t + A\varepsilon(t)\pi\right). \qquad (30)$$

The curve disentropy versus $A$ for $s(t)$ is shown in Fig. 5.

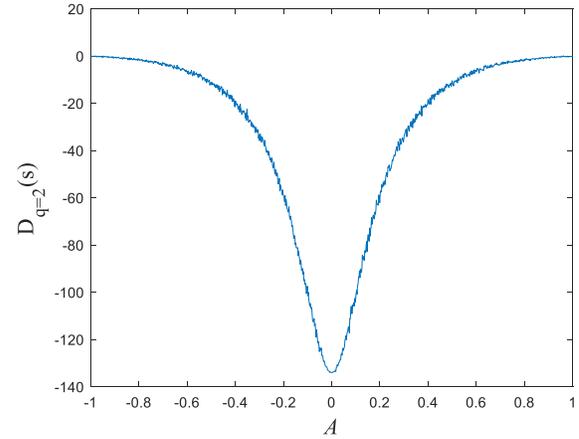

Fig. 5 – $D_2(s)$ versus $A$ for $s(t)$ given in eq. (30).

As before, the larger the noise power the larger the randomness. Comparing Figs. 3, 4 and 5, one can see that, for a telecommunication system, the noise is more harmful when it acts in the frequency or phase. In other words, coherent telecommunication systems are more susceptible to noise than telecommunication systems that use amplitude (or power) modulation.

## IV. THE DISENTROPY OF THE AUTOCORRELATION FUNCTION OF CHAOTIC SIGNALS

As it is well known, a non-linear system working in the chaotic regime can simulate a random variable, in the sense that the chaotic signal, even being deterministic, it can pass in some statistical tests of randomness. Hence, it is interesting to analyse the randomness of a chaotic signal using the disentropy of the autocorrelation. The non-linear system here considered is the logistic map

$$x_{n+1} = \lambda x_n(1 - x_n). \qquad (31)$$

The curve disentropy versus $\lambda$ for $x_n$ is shown in Fig. 6. The chaotic behaviour of the logistic map starts for $\lambda \approx 3.56995$, the red vertical line in Fig. 6. However, the randomness is not maximal at this point. It is maximal only for $\lambda$ close to 3.7. Therefore, different chaotic regimes can show different randomness. In this case, if the logistic map is used in a chaotic cryptographic scheme, it will be more secure if $\lambda \geq 3.7$.

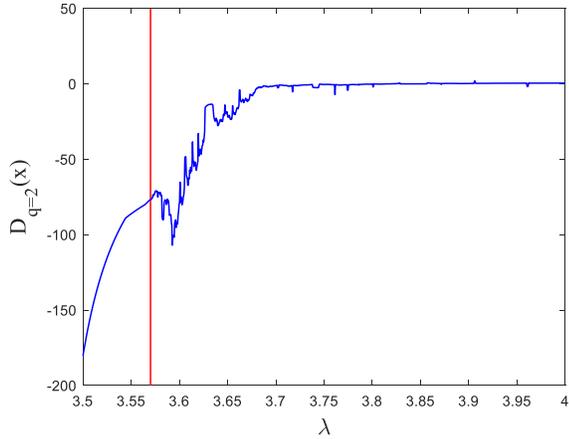

Fig. 6 – $D_2(x)$ versus $\lambda$ for $x$ given in eq. (31).

At last, consider the noisy logistic map shown in eq. (32).

$$x_{n+1} = k3.9x_n(1-x_n) + \sqrt{1-k^2}\,0.2\varepsilon_n. \tag{32}$$

The curve disentropy versus $k$ for $x_n$ is shown in Fig. 7.

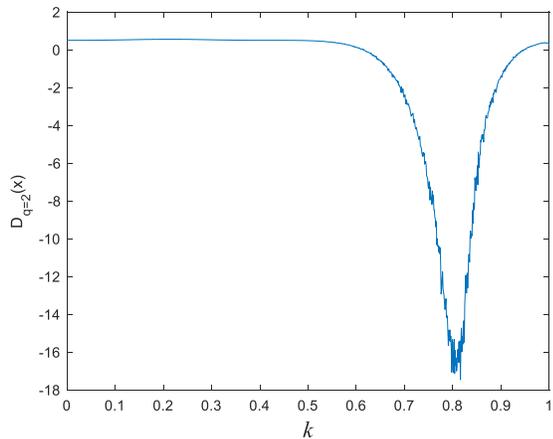

Fig. 7 – $D_2(x)$ versus $k$ for $x$ given in eq. (32).

Equation (32) is a mixture of a chaotic process with a random process hence, a large amount of randomness is expected. In fact, the minimal value of randomness, that occurs for $k \sim 0.8$, is much larger than the randomness measured when the logistic map is still working in the periodic regime ($\lambda \leq 3.56995$ in Fig. 6).

## V. CONCLUSIONS

In this work it was shown the disentropy of the autocorrelation function can be used as a measure of randomness of signals and time series. To quantify the randomness is important because one can, for example, to choose the best value of a physical parameter of the system considered aiming to decrease or increase the randomness (like in optical chaotic cryptographic schemes) or to decide what kind of medicine should be introduced to a patient according to the randomness, for instance, of the blood pressure data.

rubens.ramos@ufc.br - Department of Teleinformatic Engineering – Federal University of Ceara - DETI/UFC, C.P. 6007 – Campus do Pici - 60455-970 Fortaleza-Ce Brazil.

This work was supported by the Brazilian agency CNPq via Grant no. 307184/2018-8. Also, this work was performed as part of the Brazilian National Institute of Science and Technology for Quantum Information.